\documentstyle[12pt]{article}
\begin{document}
\title{
\begin{flushright}
{\small SMI-28-95}
\end{flushright}
\vspace{0.5cm}
Fermi-Bose duality via extra dimension.}
\author{ A.A.Slavnov \thanks{E-mail:$~~$ slavnov@class.mian.su} \\ Steklov
Mathematical Institute, Russian Academy of Sciences,\\ Vavilov st.42,
GSP-1,117966, Moscow, Russia } \maketitle

\begin{abstract}

Representation of a $D$-dimensional fermion determinant as a path
integral of exponent of a $(D+1)$-dimensional Hermitean bosonic action is
constructed.
\end{abstract}

\section {Introduction}

In the recent paper  \cite{AS} we proposed a method which allows to
present a four dimensional fermion determinant as a path integral of
exponent of a five dimensional constrained bosonic action. This
method provides a path integral bosonization of fermion models in
dimensions $D \geq2$. The problem of bosonization of fermion models in
 dimensions $D>2$ was studied previously by several authors (see for
 example \cite{Sem}, \cite{ML1}, \cite{M}, \cite{Z}, \cite{Q}), however no
 quite satisfactory solution was known. Bosonic representation is
 particulary important for lattice gauge theories like lattice QCD as it
 allows to avoid complications related to integration over grassmanian
 variables.  One way to handle the problem was proposed in papers
 \cite{ML}, \cite{ML2}, where the algorithm for approximate inversion of
 the QCD fermion determinant was formulated replacing it by an ifinite
 series of bosonic determinants.  In our paper \cite{AS} the exact
 effective bosonic action was explicitely constructed and it was proven
that when the model is considered on the finite Euclidean lattice the path
integral is convergent. Any gauge invariant observable in QCD may be
calculated in this approach.  However numerical simulations by Monte-Carlo
method are not straightforward as the effective bosonic action is not
Hermitean.

In the present paper I propose an alternative procedure which provides the
representation of a $D$-dimensional fermion
determinant in terms of a path integral of $(D+1)$-dimensional
Hermitean bosonic action. This method seems to be more
appropriate for Monte-Carlo simulations. The construction can be inverted
providing the representation of bosonic determinant in terms of a path
integral of fermionic effective action.

In the next section I illustrate the idea by the model on an
infinite lattice.  In the last section a corresponding construction for  a
 finite lattice is presented.

\section {Infinite lattice model}

To illustrate the idea I firstly present a formal construction for the
infinite lattice theory. In this formal proof I ignore the problem
of convergence of the path integrals. In the next section a rigorous proof
will be given for the case of a finite lattice.

Let $\det(B)$ be the determinant of some positive bounded Hermitean
operator $B$.
\begin{equation}
\det(B)= \int \exp \{- a^D \sum_x[ \bar{ \psi}_m(x)B^{mn} \psi_n(x)]
\}d \bar{\psi} d \psi \label{1} \end{equation} Here $ \psi_m(x)$ is a
fermion field, and $m$ is a collective index which numerates spinorial,
 colour components e.t.c. These fields are defined on a $D$-dimensional
 Euclidean lattice with the lattice spacing $a$. Our goal is to construct
 a representation for the determinant of the operator $B$ as a path
 integral of the exponent of a Hermitean bosonic action.

 Let us introduce
 bosonic fields $ \phi(x,t)$ which depend on one extra space coordinate
 $t$ and have the same spinorial and internal structure as the fields $
 \psi$. The extra coordinate $t$ is defined on the infinite one
 dimensional lattice with the lattice spacing $b$:  $t=nb, \quad - \infty
 <n< + \infty$. The fields $ \phi$ are described by a local Hermitean
 action \begin{equation} S=a^Db \sum_{n=- \infty}^{+ \infty} \sum_x [i
 \frac{ \phi^*_{n+1}(x)e^{iBb} \phi_n(x)- \phi^*_n(x)e^{-iBb
} \phi_{n+1}(x)}{2b} \label{2} \end{equation} We shall prove
the following identity \begin{equation} \det(B)=
 \lim_{b \rightarrow 0} \int \exp \{S+i( \phi^*_n(x)
\chi(x)+ \chi^*(x) \phi_n(x))] \}d \phi^*_nd \phi_nd \chi^*d \chi
\label{3}
\end{equation}
 Here $ \chi(x)$ are $D$-dimensional bosonic fields which play the role of
Lagrange multipliers imposing the constraints \begin{equation} \sum_n
\phi_n(x)= \sum_n \phi^*_n(x)=0.  \label{4} \end{equation} R.h.s.  of eq.
(\ref{3}) can be written in terms of eigenvectors of the operator $
B$, $B_{\alpha}$ being the corresponding eigenvalues
\begin{equation} I=\lim_{b \rightarrow 0} \int \exp \{b \sum_{n=-
\infty}^{+ \infty} \sum_{\alpha} [i \frac{ \phi^{*
\alpha}_{n+1}e^{iB^{\alpha}b} \phi^{\alpha}_n- \phi^{*
\alpha}_ne^{-iB^{\alpha}b} \phi^{\alpha}_{n+1}}{2b} + \label{5}
\end{equation} $$ + i( \phi^{* \alpha}_n \chi^{\alpha}+ \chi^{* \alpha}
\phi^{\alpha}_n)] \}d \phi^{* \alpha}_nd \phi^{\alpha}_nd \chi^{* \alpha}d
\chi^{\alpha} .  $$ To calculate the integral (\ref{5}) we make the
following change of variables:  \begin{equation} \phi_n^{\alpha}
\rightarrow \exp \{iB^{\alpha}nb \}\phi_n^{\alpha}, \quad
\phi_n^{\alpha*} \rightarrow \exp \{-iB^{\alpha}nb \} \phi_n^{\alpha*}
\label{6} \end{equation} After this transformation the integral (\ref{5})
written in terms of Fourier components \begin{equation} \tilde{\phi}_k=b
\sum_{n=- \infty}^{+ \infty} \phi_n \exp \{-iknb \} \label{7}
\end{equation} looks as follows \begin{eqnarray} I= \lim_{b \rightarrow 0}
\int \exp \{-(2 \pi)^{-1} \sum_{\alpha} \int_{- \frac{ \pi}{b}}^{+ \frac{
\pi}{b}}dk[| \tilde{\phi}_k|^2  \sin(kb)b^{-1} + \nonumber \\+ i(
\tilde{\phi}_k^{* \alpha} \chi^{\alpha} + \chi^{* \alpha} \tilde{
\phi}_k^{ \alpha}) \delta(k-B_{\alpha})] \}d \tilde{\phi}_k^{* \alpha} d
\tilde{\phi}_k^{\alpha} d \chi^{* \alpha} d \chi^{\alpha} \label{8}
\end{eqnarray} Integrating over $ \phi_{\alpha}$ one gets \begin{equation}
I= \lim_{b \rightarrow 0}Z \int \exp \{ -\frac{1}{2 \pi} \sum_{\alpha}
\chi^*_{\alpha}[ \sin(B_{\alpha}b)b^{-1}]^{-1} \chi_{\alpha} \}d
\chi^*_{\alpha} d \chi_{\alpha} \label{9} \end{equation} where $Z$ is the
$B_{\alpha}$ independent (infinite) constant \begin{equation} Z= \int \exp
\{ -\frac{1}{2 \pi} \sum_{\alpha} \int_{- \frac{\pi}{b}}^{
\frac{\pi}{b}}dk| \tilde{ \phi}_{\alpha}(k)|^2 \sin(kb)b^{-1} \}d
\tilde{\phi}^*_{\alpha}(k) d \tilde{\phi}_{\alpha}(k) \label{10}
\end{equation} This constant may be eliminated by a proper normalization.

If the eigenvalues $B_{\alpha}$ are limited we can choose b sufficiently
small, so that $B_{\alpha}b<<1$.
 Therefore we can replace $ \sin(B_{\alpha}b)$ by $B_{\alpha}b$, and
integrating over $ \chi_{\alpha}$ to get \begin{equation} I=
\prod_{\alpha}B_{\alpha}= \det(B) \label{11} \end{equation} The
eq.(\ref{3}) is proven. The determinant of the positive bounded
operator $B$ is presented as the path
integral of the purely bosonic Hermitean action.

However the derivation in this section was formal as we did not study the
problem of convergence. In the next section I shall present an analogous
construction for a finite lattice, where we are able to prove the
convergence of all the integrals and make a representation like eq.(\ref{3})
 quite rigorous.

\section{Finite lattice models. Lattice QCD}

Having in mind applications to QCD in this section we consider two
fermion flavours interacting vectorially with the Yang-Mills fields
$U_{\mu}$. The reason to consider two degenerate flavours is the
positivity of the square of the gauge covariant Dirac operator. It allows
to prove the convergence of the path integral for the model defined on the
finite lattice. Formally the construction goes for any number of flavours,
but a rigorous proof will be given for the case of even number of
flavours. Firstly we reduce the problem to calculation of the
determinant of Hermitean operator by using the identity \begin{equation}
\det( \hat{D}+m)= \det[\gamma_5( \hat{D}+m)], \label{12} \end{equation}
\begin{equation} \hat{D}= \frac{1}{2}\gamma_{\mu}(D^*_{\mu} -D_{\mu})
\label{13} \end{equation} In eq.(\ref{13}) $D_{\mu}$ is the lattice
covariant derivative \begin{equation} D_{\mu} \psi(x)=
\frac{1}{a}[U_{\mu}(x) \psi(x+a_{\mu})- \psi(x)] \label{14}
\end{equation} $U_{ \mu}$ is a lattice gauge field.

It is convinient to present the square of fermion determinant in the
following form
 \begin{equation}
\int \exp \{a^4 \sum_{n=1}^2 \sum_x \bar{\psi}_n(x)(
\hat{D}+m) \psi_n(x) \}d \bar{ \psi}d \psi = \det[\gamma_5( \hat{D}+m)]^2=
\label{15} \end{equation}
$$
= \int \exp \{a^4 \sum_x \bar{\psi}(x)(
\hat{D}^2-m^2) \psi(x) \} d \bar{ \psi}d \psi
$$
The operator $- \hat{D}^2+m^2$  is
the square of Hermitean operator therefore all it's eigenvalues  are positive .

We again introduce five dimensional bosonic fields $ \phi(x,t)$with the
same spinorial and internal structure as $ \psi(x)$.  The spatial
components $x$ are defined as above.  The fifth component $t$ to be
defined on the one dimensional lattice of the length $L$ with the lattice
spacing $b$:  \begin{equation} L=2Nb, \quad -N<n \leq N \label{16}
\end{equation} We choose $b<<a$.   The free
 boundary conditions in $t$ are assumed:
\begin{equation}
 \phi_n=0, \quad n \leq -N, \quad n>N  \label{18}
\end{equation}

The folowing identity is valid
\begin{equation} \int \exp \{a^4 \sum_x \bar{ \psi}(x)( \hat{D}^2-m^2)
\psi(x) \}d \bar{\psi}d \psi = \label{19} \end{equation}
$$
= \lim_{L \rightarrow \infty, b \rightarrow 0} \int \exp \{a^4b
\sum_{n=-N+1}^N \sum_x [(2b^2)^{-1}( \phi^*_{n+1}(x) \exp \{i
\gamma_5( \hat{D}+m)b \} \phi_n(x)+
$$
$$
+\phi^*_n(x) \exp \{-i \gamma_5(\hat{D}+m)b \} \phi_{n+1}(x)-2 \phi^*_n
\phi_n) + $$ $$ +\frac{i}{ \sqrt{L}}( \phi^*_n(x) \chi(x)+
\chi^*(x) \phi_n(x))] \}d \phi^*_nd \phi_nd \chi^*d \chi .  $$ R.h.s. of
eq.  (\ref{19}) can be written in terms of eigenvectors of the
operator $ \gamma_5( \hat{D}+m)$, which are denoted as $D_{\alpha}$:
\begin{equation} I=\lim_{L \rightarrow \infty, b \rightarrow 0} \int \exp
\{b \sum_{n=-N+1}^N \sum_{\alpha} [ \frac{ \phi^{*
\alpha}_{n+1}e^{iD^{\alpha}b} \phi^{\alpha}_n+ \phi^{*
\alpha}_ne^{-iD^{\alpha}b} \phi^{\alpha}_{n+1}-2 \phi^{* \alpha}
\phi^{\alpha}}{2b^2} + \label{20} \end{equation} $$ + \frac{i}{ \sqrt{
L}}( \phi^{* \alpha}_n \chi^{\alpha}+ \chi^{* \alpha} \phi^{\alpha}_n)]
\}d \phi^{* \alpha}_nd \phi^{\alpha}_nd \chi^{* \alpha}d \chi^{\alpha} .
$$ To check the convergence of the integral ( \ref{20}) we shall write it
in terms of Fourier components. For this purpose it is convinient to
replace the model on the lattice with $2N$ points and free boundary
conditions by the equivalent model on the lattice with $2N+1$ points,
periodic boundary conditions and the additional constraint
\begin{equation}
 \phi_{-N}= \phi^*_{-N}=0 \label{21}
\end{equation}
These constraints may be imposed by adding to the action the corresponding
terms with Lagrange multipliers $ \lambda$. Then the integral (\ref{20})
 may be written in terms of Fourier components \begin{equation}
 \tilde{\phi}_k= \sum_{-N}^{+N} \phi_n \exp \{ \frac{2i \pi kn}{2N+1} \}
\label{22}
\end{equation}
as follows
\begin{equation}
I= \lim_{L \rightarrow \infty, b  \rightarrow 0} \int \exp \{b(2N+1)^{-1}
\sum_{\alpha}[ \sum_{k=-N}^{N}| \tilde{\phi}^{\alpha}_{k}|^2[ \cos(
\frac{2 \pi k}{2N+1}-D^{\alpha}b)-1](2b^2)^{-1} + \label{23}
\end{equation} $$ +ib(L)^{- \frac{1}{2}}( \tilde{ \phi}^{* \alpha}_0
\chi^{\alpha}+ \chi^{* \alpha} \tilde{ \phi}^{\alpha}_0)+ib(2N+1)^{-1}
 \sum_k( \tilde{ \phi}^{* \alpha}_k \exp \{ \frac{2i \pi kN}{2N+1} \}
\lambda_{\alpha} + $$ $$ +\lambda^*_{\alpha} \tilde{ \phi}^{\alpha}_k \exp
\{ \frac{-2i \pi kN}{2N+1} \})] \} d \tilde{ \phi}^*_{\alpha}(k) d
\tilde{\phi}_{\alpha}(k) d \chi^*_{\alpha} d \chi_{\alpha} d
\lambda^*_{\alpha} d \lambda_{\alpha} $$ The representation (\ref{23})
makes the convergence of the integral obvious.

Now we come back to coordinate space and to calculate the integral
(\ref{20}) we make the following change of variables:  \begin{equation}
\phi_n^{\alpha} \rightarrow \exp \{iD^{\alpha}nb \}\phi_n^{\alpha}, \quad
\phi_n^{\alpha*} \rightarrow \exp \{-iD^{\alpha}nb \} \phi_n^{\alpha*}
\label{24} \end{equation} $$ \lambda^{\alpha} \rightarrow \exp
\{iD^{\alpha}Nb \} \lambda^{\alpha}, \quad \lambda^{* \alpha} \rightarrow
\exp \{-iD^{\alpha}Nb \} \lambda^{* \alpha} $$ Then one gets
\begin{equation} I= \lim_{L \rightarrow \infty, b \rightarrow 0} \int \exp
\{b \sum_{n=-N}^{+N}[ \frac{ \phi^{* \alpha}_{n+1} \phi^{\alpha}_{n}+
 \phi^{* \alpha}_n \phi^{ \alpha}_{n+1} -2 \phi^{* \alpha}_n \phi^{
 \alpha}_n}{2b^2} + \label{25} \end{equation} $$ +i(L)^{- \frac{1}{2}}(
 \phi^{* \alpha}_ne^{iD^{\alpha}nb} \chi^{\alpha}+ \chi^{*
 \alpha}e^{-iD^{\alpha}nb} \phi^{\alpha}_n)] + $$ $$ +i( \phi^{*
\alpha}_{-N} \lambda^{ \alpha} + \lambda^{* \alpha} \phi^{\alpha}_{-N}) \}
 d \phi^{* \alpha}_nd \phi^{\alpha}_n d \chi^{* \alpha} d \chi^{\alpha} d
\lambda^{* \alpha} d \lambda^{\alpha} $$ Now the quadratic form in the
exponent does not depend on $D_{\alpha}$ and therefore the corresponding
determinant is a trivial constant. So we can calculate the integral by
finding the stationary point of the exponent, which is defined by the
following equations:  $$ b^{-2}( \phi^{* \alpha}_{n+1}+ \phi^{*
\alpha}_{n-1}-2 \phi^{* \alpha}_n)+iL^{- \frac{1}{2}} \chi^{* \alpha}
e^{-iD_{\alpha}nb}=0, \quad n \neq -N $$ \begin{equation} b^{-2}( \phi^{
\alpha}_{n+1}+ \phi^{ \alpha}_{n-1}-2 \phi^{ \alpha}_n)+iL^{- \frac{1}{2}}
\chi^{ \alpha} e^{iD_{\alpha}nb}=0, \quad n \neq -N \label{26}
\end{equation} $$ b^{-2}( \phi^{* \alpha}_{-N+1}-2 \phi^{*
\alpha}_{-N})+iL^{- \frac{1}{2}} \chi^{* \alpha} e^{iD_{\alpha}Nb}+i
\lambda^{* \alpha}=0 $$  $$ b^{-2}( \phi^{ \alpha}_{-N+1}-2 \phi^{
 \alpha}_{-N})+iL^{- \frac{1}{2}} \chi^{ \alpha} e^{-iD_{\alpha}Nb}+i
\lambda^{\alpha}=0 $$ $$ \phi_{-N}= \phi^{*}_{-N}=0, \quad \phi_{N+1}=
 \phi^{*}_{N+1}=0 $$ For small $b$ these equations can be approximated by
 the differential equations:  \begin{equation} \ddot \phi^{* \alpha}
+iL^{- \frac{1}{2}} \chi^{* \alpha} e^{-iD^{\alpha}t}=0\label{27}
 \end{equation} $$ \ddot \phi^{ \alpha} +iL^{- \frac{1}{2}} \chi^{ \alpha}
 e^{iD^{\alpha}t}=0 $$ $$ \phi( \frac{L}{2})= \phi( -\frac{L}{2})=0, \quad
 \phi^*( \frac{L}{2})= \phi^*(- \frac{L}{2})=0 $$ The solution of these
eq.s is \begin{equation} \phi^{* \alpha}= \frac{i \chi^{* \alpha}}{
\sqrt{L}(D^{ \alpha})^2} e^{-iD^{\alpha}t}- \label{28} \end{equation} $$ -
\frac{ \chi^{* \alpha}}{(D^{ \alpha})^2 \sqrt{L}}[ \frac{2t}{L} \sin(
\frac{D^{\alpha}L}{2})+i \cos ( \frac{D^{\alpha}L}{2})] $$ $$ \phi^{
\alpha}= \frac{i \chi^{\alpha}}{ \sqrt{L}(D^{ \alpha})^2}
e^{iD^{\alpha}t}+ $$ $$ + \frac{ \chi^{ \alpha}}{(D^{ \alpha})^2
\sqrt{L}}[ \frac{2t}{L} \sin( \frac{D^{\alpha}L}{2})-i \cos (
\frac{D^{\alpha}L}{2})] $$ Substituting this solution to the eq.(
\ref{25}) we get \begin{equation} I= \lim_{L \rightarrow \infty} \int \exp
\{- \frac{- \chi^{* \alpha} \chi^{\alpha}}{(D^{\alpha})^2}+O(L^{-1}) \}=
\prod_{\alpha}(D^{\alpha})^2= \det(-D^2+m^2) \label{29} \end{equation} The
 eq.(\ref{19}) is proven. It is clear from eqs.(\ref{28},\ref{29}), that
in the limit $L \rightarrow \infty$ the dependence on the boundary
conditions disappear.

For practical calculations it may be useful to linearize the
 eq.(\ref{19}). It can be done in a straightforward way using the fact
that the operator $ \gamma_5( \hat{D}+m)$ is bounded \cite{ML}:
\begin{equation} ||\gamma_5(\hat{D}+m)|| \leq 8a^{-1}+m \label{30}
\end{equation}. Therefore for any $ \alpha$ the product
$D^{\alpha}b<<1$. The exponent $ \exp \{ibD_{\alpha} \}$ may be replaced
by $1+ibD_{\alpha}- \frac{1}{2}b^2(D_{\alpha})^2$ and linearized version
of the eq.(\ref{19}) looks as follows \begin{equation} \int \exp \{a^4
\sum_x \bar{ \psi}(x)( \hat{D}^2-m^2) \psi(x) \}d \bar{\psi}d \psi =
\label{31} \end{equation} $$ = \lim_{L \rightarrow \infty, b \rightarrow
0} \int \exp \{a^4b \sum_{n=-N+1}^N \sum_x [(2b^2)^{-1}( \phi^*_{n+1}(x)
\phi_n(x)+ \phi^*_n(x) \phi_{n+1}(x)-2 \phi^*_n \phi_n) + $$ $$ +i[
\phi^{*}_{n+1}(x) \gamma_5( \hat{D}+m) \phi_n(x)- \phi^{*}_n(x) \gamma_5
 (\hat{D}+m) \phi_{n+1}(x)][2b]^{-1}+ $$ $$ +\frac{1}{2}
\phi^{*}_n(D^2-m^2) \phi_n +\frac{i}{ \sqrt{L}}( \phi^*_n(x) \chi(x)+
\chi^*(x) \phi_n(x))] \}d \phi^*_nd \phi_nd \chi^*d \chi .  $$ One can get
 rid off the last term in the exponent by solving explicitely the
constraints \begin{equation} \sum_{n=-N+1}^N \phi_n(x)=0 \label{32}
\end{equation} Substituting the solution of this equation to the integral
(\ref{31}) we get a representation for the determinant of the square of
the gauge covariant Dirac operator as the path integral of the exponent of
the purely bosonic Hermitean action.

\section {Discussion}

Comparing the results presented above with the previous ones \cite{AS} we
note that the new bosonic representation for gauge covariant Dirac
operator is based on the Hermitean action and therefore is better suited
for numerical simulations. We also got rid off the extra parameter $
\lambda$ which was present in our first paper. One can hope that the
construction given in this paper will allow to use directly the
Monte-Carlo method for calculation of fermion determinant.

{\bf Acknowledgements.} \\
This researsh was supported in part by Russian Basic Research
Fund under grant 94-01-00300a.$$ ~ $$ \begin{thebibliography}{99} {\small
\bibitem{AS}
A.A.Slavnov {\it Bosonization of fermion determinants} hep-th 9507152, to
appear in Phys.Lett.B.  \bibitem{Sem}
G.W.Semenoff, Phys.Rev.Lett.61 (1988) 817.  \bibitem{ML1} M.L$\ddot
u$sher, Nucl.Phys. B326 (1989) 557. \bibitem{M} E.C.Marino, Phys.Lett.
B263 (1991) 63. \bibitem{Z} L.Huerta,F.Zanelli, Phys.Rev.Lett.  71 (1993)
3622.  \bibitem{Q} C.P.Burgess, C.A.L$\ddot u$tken, F.Quevedo, Phys.Lett.
B336 (1994) 18. \bibitem{ML} M.L$\ddot u$sher Nucl.Phys.  B418 (1994)
637.  \bibitem{ML2} B.Bunk, K.Jansen, B.Jegerlehner, M.L$\ddot u$sher,
H.Simma, R.Sommer, Nucl.Phys.B (Proc.Suppl.)42 (1995) 49.  } \end
{thebibliography} \end{document}